\documentclass{desyproc}

\newcommand{\MSSM}{\textmd{\textsc{mssm}}}
\newcommand{\Mh}{\ensuremath{M_h}}
\newcommand{\SUSY}{{\sc susy}}

\newcommand{\SUSYQCD}{{\sc susy-qcd}}
\newcommand{\LHC}{{\sc lhc}}

\newcommand{\Mt}{\ensuremath{m_t}}
\newcommand{\Mg}{\ensuremath{m_{\tilde{g}}}}
\newcommand{\Mst}{\ensuremath{m_{\tilde{t}_{1,2}}}}
\newcommand{\Mq}{\ensuremath{m_{\tilde{q}}}}
\newcommand{\hthreel}{{\sc H3m}}
\newcommand{\order}[1]{\ensuremath{{\cal O}\left( {#1} \right)}}

\title{%
  \vskip-4\baselineskip%
  {\normalsize HU-EP-10/40\\%
    \normalsize SFB/CPP-10-71}%
  \vskip-3\baselineskip%
  \vskip4\baselineskip
Three-Loop Predictions for the Light Higgs Mass in the \MSSM}
\author{{\slshape Philipp Kant}\\[1ex]
  Humboldt-Universit\"at zu Berlin, Newtonstr.~15, 12489~Berlin}
\contribID{xy}  
\confID{1964}
\desyproc{DESY-PROC-2010-01}
\acronym{PLHC2010}
\doi            

\begin{document}
\maketitle

\begin{abstract}
  The Minimal Supersymmetric Extension of the Standard Model (\MSSM{})
  features a light Higgs boson, the mass \Mh{} of which is
  predicted by the theory.  
  Given that the \LHC{} will be able
  to measure the mass of a light Higgs with great accuracy, a precise
  theoretical calculation of \Mh{} yields an important test of the
  \MSSM{}.
  In order to deliver this precision, we present three-loop radiative
  corrections of \order{\alpha_t\alpha_s^2} and provide a computer
  code that combines our results with corrections to \Mh{} at lower
  loop orders that are available in the literature.

\end{abstract}

\section{Introduction}
The Higgs sector of the Minimal Supersymmetric Extension of the
Standard Model (\MSSM) consists of a two-Higgs doublet model, which is
tightly constrained by Supersymmetry.  In particular, the quartic
terms of the Higgs potential are completely fixed by the gauge
couplings.  Thus, it is possible to describe the \MSSM{} Higgs sector
through only two new (with respect to the Standard Model) parameters,
which are usually taken to be the mass $M_A$ of the pseudoscalar Higgs
and the ratio $\tan\beta=\tfrac{v_2}{v_1}$ of the vacuum
expectation values of the Higgs doublets.
In particular, \Mh{}, the mass of the light scalar Higgs boson, can be
predicted, and at the tree-level only these two parameters enter the
prediction, leading to an upper bound of $\Mh\leq M_Z$.  However,
\Mh{} is sensitive to virtual corrections to the Higgs propagator that
shift this upper bound significantly.  These virtual corrections
depend on all the Supersymmetry breaking parameters.
This sensitivity to virtual corrections, combined with the
great precision with which the Large Hadron Collider (\LHC) will be
able to measure the mass of a light Higgs, allows \Mh{} to be used as
a precision observable to test supersymmetric models -- assuming that
the theoretical uncertainties are sufficiently small and under control.

Consequently, the one- and two-loop corrections to \Mh{} have been
studied extensively in the literature (see, for example~\cite{%
  Hempfling:1994qq, Zhang:1998bm, Heinemeyer:1998jw,%
  Pilaftsis:1999qt,Carena:2000dp,Espinosa:2001mm,%
  Degrassi:2001yf,Martin:2002wn}).  The remaining uncertainty has been
estimated to be about
$3-5\,$GeV~\cite{Degrassi:2002fi,Allanach:2004rh}.  Recently, also
three-loop corrections have become available.  The leading- and
next-to-leading terms in $\ln(M_{SUSY}/M_t)$, where $M_{SUSY}$ is the
typical scale of \SUSY{} particle masses, have been obtained
in~\cite{Martin:2007pg}.  Motivated by the observation that the
contributions from loops of top quarks and their superpartners, the
stops, are dominant at the one- and two-loop level, we have calculated
three-loop \SUSYQCD{} corrections to these diagrams.  These
corrections are of \order{\alpha_t\alpha_s^2}, where $\alpha_t$ is the
coupling of the Higgs to the top quarks.  A first result has been
obtained in~\cite{Harlander:2008ju}.  There, we assumed that all the
superpartners had approximately the same mass.  This restriction has
been dropped recently in~\cite{Kant:2010tf}.

\section{Organisation of the Calculation}
A major difficulty in obtaining the results of~\cite{Kant:2010tf} was
the presence of many different mass scales -- the masses \Mt{} of the
top quark, \Mg{} of the gluino, \Mst{} of the stops and \Mq{} of the
partners of the light quarks -- in the three-loop propagator diagrams.
Assuming that there is a distinct hierarchy between these masses, they
can be disentangled by the method of asymptotic
expansions~\cite{Smirnov:2002pj}, yielding an expansion of the
diagrams in small mass ratios and logarithms of mass ratios.  Working
in the effective potential approximation, we set the external momentum
flowing through the Higgs propagator to zero and are left with tadpole
integrals with a single mass scale, which are known and implemented in
the {\sc form}~\cite{Vermaseren:2000nd} program {\sc
  matad}~\cite{Steinhauser:2000ry}.

However, as the masses of the superpartners are not known, it is not
clear which hierarchy one should assume.  We solve this by computing
the diagrams for many different hierarchies.  Then, when given a point
in the \MSSM{} parameter space, we choose whichever hierarchy fits
best and evaluate \Mh{} using the calculation in the chosen hierarchy.
To choose the best hierarchy and to estimate the error introduced by
the asymptotic expansion, we compare, at the two-loop level, our
expanded result with the result of~\cite{Degrassi:2001yf}, which
contains the full mass dependence.

\begin{figure}
    \begin{center}%
      \includegraphics[width=.5\textwidth]{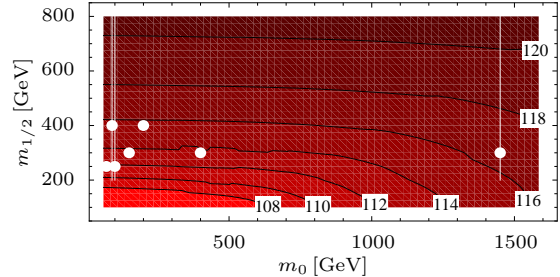}%
    \end{center}%
    \vskip-\baselineskip
  \caption{Prediction for the value of $M_h$ (in GeV)
    for {\sc msugra} scenario with $\tan\beta=10$, $A_0=0$, as
    evaluated by \hthreel{}.  The white lines and points
    indicate the benchmark scenarios of~\cite{Allanach:2002nj}.}
  \label{fig:mh}
\end{figure}

For convenience, we have written the {\sc Mathematica} package
{\hthreel}~\cite{h3m}, which automatically performs the choice of the
best fitting hierarchy and provides a {\sc susy Les Houches} interface
to our calculation.  This allows to perform parameter scans as in
Fig.~\ref{fig:mh}.  In order to get a state-of-the art prediction for
\Mh{}, we include all available contributions to \Mh{} at the one- and
two-loop level that are implemented in {\sc
  FeynHiggs}~\cite{FeynHiggs}.  
For details on the usage and inner workings of the program,
we refer to~\cite{Kant:2010tf}.

\section{Estimating the Theoretical Uncertainty}
We observe that the dependence of \Mh{} on the renormalisation
prescription, which is often used as a guesstimate for the uncertainty
due to unknown higher order corrections, reduces drastically when one
goes from two to three loops.
But since we also find that the size of the three-loop corrections can
be of the order of one to two GeV, which is rather large given that
the two-loop corrections are only about a factor of two larger, we
prefer to be conservative in our estimation of the theoretical
uncertainty.  Assuming a geometric progression of the perturbative
series, we get for {\sc msugra} scenarios an uncertainty due to
missing higher order corrections of $100\,$MeV to $1\,$GeV, depending
on the value of $m_{1/2}$.  The parametric uncertainty due to
$\alpha_s$, \Mt{} and \Mst{} is of the same order of magnitude.  The
uncertainty introduced by the expansion in mass ratios amounts to at
most $100\,$MeV~\cite{Kant:2010tf}.

\section{Conclusions}
We present a calculation of the \order{\alpha_t\alpha_s^2} corrections
to \Mh{}, shifting the value of \Mh{} by about $1\,$GeV. We
provide a computer code combining our results with corrections from
lower loop orders, thus enabling a state-of-the-art prediction of
\Mh{}.  Our calculation lowers the theoretical uncertainty due to
missing higher orders to the same magnitude as the parametric
uncertainty.

This work was supported by the DFG through SFB/TR~9 and by the
Helmholtz Alliance ``Physics at the Terascale''.

\section{Bibliography}

\begin{footnotesize}

\end{footnotesize}

\end{document}